\documentclass[superscriptaddress,twocolumn,epsf,prb,amsmath,amssymb,showpacs]{revtex4-1}

\pdfoutput=1

\usepackage{graphicx}
\usepackage{url}
\usepackage{amsmath}
\usepackage{amssymb}
\usepackage{xcolor}
\usepackage[T1]{fontenc}

\definecolor{darkblue}{rgb}{0.,0.,0.4}
\usepackage[pdftex,colorlinks=true,linkcolor=blue,citecolor=blue,urlcolor=blue,breaklinks]{hyperref}

\begin{document}

\title{An electric-field driven Mott metal-insulator transition in correlated thin films: an inhomogeneous dynamical mean-field theory approach}
\author{P. Bakalov}
\email{petar.bakalov@fys.kuleuven.be}
\affiliation{Departement Natuurkunde en Sterrenkunde, KULeuven, Celestijnenlaan 200D, B-3001 Leuven, Belgium}
\author{D. Nasr Esfahani}
\email{Davoud.NasrEsfahani@uantwerpen.be}
\affiliation{Departement Fysica, Universiteit Antwerpen, Groenenborgerlaan 171, B-2020 Antwerpen, Belgium}
\author{L. Covaci}
\email{lucian@covaci.org}
\affiliation{Departement Fysica, Universiteit Antwerpen, Groenenborgerlaan 171, B-2020 Antwerpen, Belgium}
\author{F. M. Peeters}
\email{Francois.Peeters@uantwerpen.be}
\affiliation{Departement Fysica, Universiteit Antwerpen, Groenenborgerlaan 171, B-2020 Antwerpen, Belgium}
\author{J. Tempere}
\email{jacques.tempere@uantwerpen.be}
\affiliation{Departement Fysica, Universiteit Antwerpen, Universiteitsplein 1, B-2610 Wilrijk, Belgium}
\author{J.-P. Locquet}
\email{jeanpierre.locquet@fys.kuleuven.be}
\affiliation{Departement Natuurkunde en Sterrenkunde, KULeuven, Celestijnenlaan 200D, B-3001 Leuven, Belgium}

\begin{abstract}
Simulations are carried out based on the dynamical mean-field theory (DMFT) in order to investigate the properties of correlated thin films for various values of the chemical potential, temperature, interaction strength, and applied transverse electric field. Application of a sufficiently strong field to a thin film at half-filling leads to the appearance of conducting regions near the surfaces of the film, whereas in doped slabs the application of a field leads to a conductivity enhancement on one side of the film and a gradual transition to the insulating state on the opposite side. In addition to the inhomogeneous DMFT, an independent layer approximation (ILA) is considered, in which the properties of each layer are approximated by a homogeneous bulk environment. A comparison between the two approaches reveals that the less expensive ILA results are in good agreement with the DMFT approach, except close to the metal-to-insulator transition points and in the layers immediately at the film surfaces. The hysteretic behavior (memory effect) characteristic of the bulk doping driven Mott transition persists in the slab.
\end{abstract}

\maketitle

\section{Introduction}
Due to their interesting and useful properties, strongly correlated materials have attracted the attention of both theorists\cite{PhysRevLett.101.066802,PhysRevLett.103.116402,PhysRevB.78.115103} and experimentalists\cite{KMartens2012,Jeong22032013,KMartens2014,Nakano2012,Caviglia2008} in recent decades. In particular, the discoveries of high-temperature superconductivity in cuprates and of colossal magnetoresistance in manganites spurred interest in the class of materials exhibiting the Mott metal-to-insulator transition.\cite{RevModPhys.70.1039,Limelette03102003} The existence of this correlation-driven transition in bulk transition-metal oxide systems has encouraged investigations for potential applications in electronics.\cite{Jeong22032013,KMartens2014,Nakano2012} Intriguing physics has also been observed at interfaces between strongly correlated materials;\cite{Okamoto2004} for example, a metallic or even superconducting phase appears at the interface between LaTiO$_3$ and SrTiO$_3$, which are insulating paramagnets in bulk.\cite{Hwang2012,Caviglia2008} The appearance of such interface phases can be partially understood at a qualitative level as the result of charge transfer. A detailed understanding, however, remains elusive. Due to the difficulties inherent in the theoretical treatment of strongly-correlated systems, attaining a detailed understanding may be quite challenging even when dealing with simpler bulk systems.

The exponential growth of the Hilbert space with the number of particles makes a direct numerical solution of an interacting quantum system unfeasible, except for systems with a very limited number of particles. Standard theoretical approaches such as density functional theory (DFT) are known to fail when electron-electron interactions are strong.\cite{RevModPhys.78.865} This has stimulated the development of alternative methods. The Gutzwiller approximation\cite{PhysRevLett.10.159,PhysRevB.57.6896,PhysRevB.41.9452} and slave boson techniques\cite{PhysRevLett.57.1362,PhysRevB.76.155102} have been used to treat the low-energy physics of bulk materials and - more recently - thin films\cite{PhysRevB.75.195117,PhysRevB.81.115134,PhysRevB.85.085110,PhysRevB.87.035131,0953-8984-26-7-075601,PhysRevB.90.155118} in the presence of an electric field.\cite{PhysRevB.85.085110,PhysRevB.87.035131} The slave boson approach has been extended to allow access to higher-energy excitations\cite{PhysRevB.48.11453}, while the Gutzwiller approximation only provides insight into the metallic state. Other notable approaches include the density renormalization group (DMRG), which works best for one-dimensional systems.\cite{Verstraete2014} Over the last 25 years dynamical mean-field theory (DMFT) has emerged as one of the most promising frameworks for the treatment of strongly correlated systems.\cite{RevModPhys.68.13,PhysRevLett.62.324} DMFT involves a mapping of a lattice problem, such as the Hubbard model,\cite{Hubbard26111963} onto an interacting quantum impurity model\cite{PhysRev.124.41} supplemented by a self-consistency condition. The mapping is exact in the limit of infinite dimensions. An advantage of DMFT is that it is formulated in the thermodynamic limit and that it allows one to non-perturbatively interpolate between the strong and weak coupling regimes, thereby treating different phases on an equal footing. In its simplest version, single-site DMFT, it neglects spatial correlations, but captures local temporal correlations. In combination with the local density approximation (LDA) to DFT it has provided valuable new insights into the physics of a number of strongly correlated bulk materials.\cite{RevModPhys.78.865,RevModPhys.77.1027,KHeldAdvPhys2007}

A better understanding of the metal-insulator transition in systems with lower translational symmetry (such as thin films or interfaces) and in the presence of an electric field is likely to aid experimental efforts to control the properties of such systems, which is relevant for possible applications in e.g. memory devices and high-speed electronics. Potthoff and Nolting\cite{PhysRevB.59.2549,PhysRevB.60.7834} used a spatially resolved version of DMFT (inhomogeneous DMFT, or IDMFT) to study layered systems. Similar techniques were later applied to inhomogeneous multilayered nanostructures.\cite{PhysRevLett.103.116402,PhysRevB.85.205444,PhysRevLett.101.066802,PhysRevB.79.045130} DMFT-based approaches were recently proposed for the study of "Mott \textit{p-n} junctions"\cite{PhysRevB.87.035137} and correlated capacitors.\cite{PSSA:PSSA201300418} In these approaches each layer was approximated by a two-dimensional bulk system with the appropriate value of the chemical potential, and a Poisson solver was used to study the charge redistribution in different regimes. 

In this paper we use the IDMFT to investigate the effect of an applied electric field on the properties of a slab (a stack of coupled correlated two-dimensional layers) at various temperatures and doping levels. We also approximate the properties of each layer in the slab by a three-dimensional bulk system in which the value of the chemical potential is equal to the local value of the potential in that layer. We refer to this approach as the independent layer approach (ILA) and we carry out a systematic comparison between the ILA and the full DMFT calculation. This comparison is of some interest, since the ILA is significantly faster than the full calculation. We investigate the effect of screening by considering the Coulomb interaction between planes in the slab at a mean-field level. 

The structure of this paper is as follows. We describe the model and method in Sec. \ref{mnm}. Our results are discussed in Sec. \ref{rnd}. Sec. \ref{conc} provides a summary.

\section{Model and method}
\label{mnm}
The general approach that we use is described in detail elsewhere,\cite{PhysRevB.59.2549,PhysRevB.60.7834} and we only give a brief description here. Our starting point is the single-band Hubbard Hamiltonian:
\begin{equation}
\begin{split}
H =& H_{kin} + H_U + H_{CP} \\
=& -\sum_{ij\gamma\delta\sigma} t_{\gamma\delta}^{ij} c_{i\gamma\sigma}^\dagger c_{j\delta\sigma} + U\sum_{i\gamma} n_{i\gamma\downarrow} n_{i\gamma\uparrow}\\
& - \mu \sum_{i\gamma\sigma} n_{i\gamma\sigma},
\end{split}
\end{equation}
where $c_{i\gamma\sigma}^\dagger$ ($c_{i\gamma\sigma}$) is a fermionic creation (annihilation) operator for a particle of spin $\sigma$ ($\sigma = \uparrow, \downarrow$) at site $\gamma$ in plane $i$, $n_{i\gamma\sigma}=c_{i\gamma\sigma}^\dagger c_{i\gamma\sigma}$, $t^{ij}_{\gamma\delta}$ is a hopping matrix, $U$ is the on-site Coulomb repulsion, $\mu$ is the chemical potential. We work with a simple cubic lattice and assume that only neares-neighbor hopping takes place, and that the intra- and inter-plane hopping parameters are equal, i.e. all hopping matrix elements vanish except $t^{ii}_{<\gamma\delta>} = t^{ii\pm1}_{\gamma\gamma} \equiv t$, where the angular brackets indicate that sites $\gamma$ and $\delta$ are next neighbors. We take $t = 1$, which sets the energy scale in the problem. In order to include the effect of an externally applied electric field, we include the following term:
\begin{equation}
H_{EP} = \sum_{i\gamma\sigma} v_i n_{i\gamma\sigma},
\end{equation}
where $v_i = -Vt(i/(L-1) -1/2)$ is the on-site potential for plane $i$ ($V$ is the overall potential drop across the slab and $L$ is the number of planes in the slab). This form assumes that the charge in the slab does not screen the externally applied field and neglects any Coulomb interaction between the layers themselves. We also include the effects of screening by including a correction to the bare potential of the form:\cite{PhysRevB.70.195342,PhysRevB.87.035137}
\begin{equation}
v_{Coulomb,i} = \alpha \sum_{j\ne i} (n_j - n_{bulk})|i - j|,
\label{coulomb}
\end{equation}
where $n_{bulk}$ is the number of electrons per site in the neutral solid, $n_j = n_{j,\uparrow} + n_{j,\downarrow}$ is the average number of electrons at sites in plane $j$ (site index $\gamma$ is suppressed, since density does not vary within a plane), and $\alpha$ determines the interaction strength between charges, $\alpha = \frac{e^2 d}{2\epsilon_r\epsilon_0A}$. In the last expression $e$ is the elementary charge, $\epsilon_r$ is the relative dielectric constant of the solid, $d$ is the distance between planes in our thin film, and $A$ is the area per atom within each plane. For example, taking $d = 10^{-9}m$, $A = d^2$, and $\epsilon_r = 10$, one has $\alpha \approx 2eV$. 

In the absence of screening, we use the standard DMFT loop: starting from a guess for the slab self-energy, $\hat{\Sigma}_\sigma(i \omega_n) = \text{Diag}(\Sigma_{\sigma1}(i\omega_n), \Sigma_{\sigma2}(i\omega_n), \dots)$, we obtain the bare Green's function (also referred to as the Weiss field) of the effective action for each layer in the slab, $\mathcal{G}_{0,i\sigma}(i\omega_n)$: 
\begin{equation}
\mathcal{G}^{-1}_{0,i\sigma}(i\omega_n) = [G_{ii\sigma}(i\omega_n)]^{-1}+\Sigma_{\sigma i}(i\omega_n),
\label{g0}
\end{equation}
where $\omega_n = (2n+1)\pi/\beta$ are the fermionic Matsubara frequencies at temperature $T = 1/\beta$, the index $i = 1, 2, \dots L$, and $G_{ii\sigma}$ is a diagonal element of the $k$-integrated slab Green's function corresponding to $\hat{\Sigma}_\sigma(i\omega_n)$:
\begin{equation*}
\hat{G}_{\sigma}(i\omega_n) = \frac{1}{N_k}\sum_{\mathbf{k}\in BZ}\frac{\mathbf{1}}{ (i\omega_n+ \mu)\mathbf{1} - \hat{H}_\sigma(\mathbf{k}) -\hat{\Sigma}_\sigma(i\omega_n)},
\end{equation*}
where $\mathbf{1}$ is a $L\times L$ unit matrix ($L$ is the number of layers in the slab), $N_k=N_{k_x}\times N_{k_y}$, $N_{k_x}$ and $N_{k_y}$ are number of $k$-points in the $x$ and $y$ directions of the Brillouin zone, respectively, and $\hat{H}_\sigma(\mathbf{k})$ is a tri-diagonal $L\times L$ matrix in a mixed (Fourier/real-space) basis obtained from the sum of $H_{kin}$ and $H_{EP}$:
\begin{equation*}
\hat{H}_\sigma(\mathbf{k}) = 
\begin{pmatrix}
\epsilon_\mathbf{k} +  v_1 & -t & 0 & \dots    \\
-t & \epsilon_\mathbf{k}  + v_2 & -t & \dots  \\
0 & -t &\epsilon_\mathbf{k}  + v_3 &  \dots  \\
\vdots & \vdots &  \ddots  &  \\
\end{pmatrix},
\end{equation*}
$\epsilon_\mathbf{k}=-2t(\cos{k_x}+\cos{k_y})$ being the in-plane dispersion. The impurity Green's function for each layer, $G_{imp,i\sigma}(i\omega_n)$, is obtained by solving the respective impurity problem determined by $\mathcal{G}_{0,i\sigma}$. From the impurity Green's function a new $\Sigma_{\sigma i}(i\omega_n)$ is obtained using Eq. (\ref{g0}) with $G_{ii\sigma}$ replaced by $G_{imp,i\sigma}$. $\hat{G}_{\sigma}(i\omega_n)$ is only recalculated once the solver has swept through all the layers and all $\hat{\Sigma}_{\sigma i}$ are updated. This is repeated until convergence.

To include screening, we adjust the potential according to Eq. (\ref{coulomb}) after each solver sweep (just before the recalculation of $G_{imp,i\sigma}(i\omega_n)$). For the doped calculations (not screened) it is also necessary to adjust the chemical potential of the slab between solver sweeps in order to keep the density fixed. 

To solve the resulting DMFT equations we use the hybridization expansion continuous-time quantum Monte Carlo solver provided as a part of TRIQS (a toolbox for research on interacting quantum systems)\cite{TRIQS,PhysRevLett.97.076405,PhysRevB.74.155107,PhysRevB.84.075145}. In the segment picture\cite{PhysRevB.74.155107} the on-site densities are obtained directly from the solver. We estimate the quasiparticle weight $Z$ from the imaginary frequency self-energy:\cite{PhysRevB.83.235113}
\begin{equation}
Z = \left( 1- \frac{\text{Im}\Sigma(i\omega_0)}{\omega_0}\right)^{-1}.
\end{equation}
To estimate $A(\omega=0)$, the density of states at the Fermi level, we use the relation:\cite{PhysRevB.83.235113}

\begin{equation}
\beta G(\beta/2) \xrightarrow{\beta\rightarrow \infty}-\pi A(0).
\label{dos_estimate}
\end{equation}

Aside from the full DMFT calculation described above, we also calculate the expected density distribution in the slab assuming that the density of electrons per site in each layer, $n_i$, is determined solely by the local value of the chemical potential and independently of the rest of the layers. Within this approach, we assume that
\begin{equation}
n_i = n_{3D}(\mu_i),
\label{indep_plane}
\end{equation}
where $n_{3D}(\mu)$ is the density per site of the three-dimensional Hubbard model on a cubic lattice with nearest-neighbor hopping at chemical potential $\mu$ and the same values of $U, t$ and $\beta$ as in the slab, and $\mu_i = \mu + v_i$. $n_{3D}(\mu)$ is obtained by single-site DMFT for a set of predetermined values of $\mu$. The density for values of $\mu$ that do not coincide with values in that predetermined set is obtained from a linear interpolation:
\begin{equation}
n(\mu) = n(\mu_{left})x +n(\mu_{right})(1-x),
\end{equation}
where $\mu_{left, right}$ are the two values of the chemical potential in the set which are closest to $\mu$ (such that $\mu_{left} \le \mu \le \mu_{right}$), and $x = (\mu_{right} - \mu)/(\mu_{right} - \mu_{left})$. In a similar manner we obtain ILA estimates for the density of states at the Fermi level, $A(\omega = 0)$, and the QP weight $Z$. Two stable solutions of the DMFT equations of the 3D Hubbard model, a metallic and an insulating one, exist in parts of the $U, \beta, \mu$-range which we explore. For the ILA we use the density that correpsonds to the metallic branch.

In all our calculations we enforce a paramagnetic phase, i.e. we set $\hat{\Sigma}_\uparrow=\hat{\Sigma}_\downarrow$.

\section{Results}
\label{rnd}
\subsection{Coexistence}

\begin{figure}[]
  \includegraphics*[width=.5\textwidth]{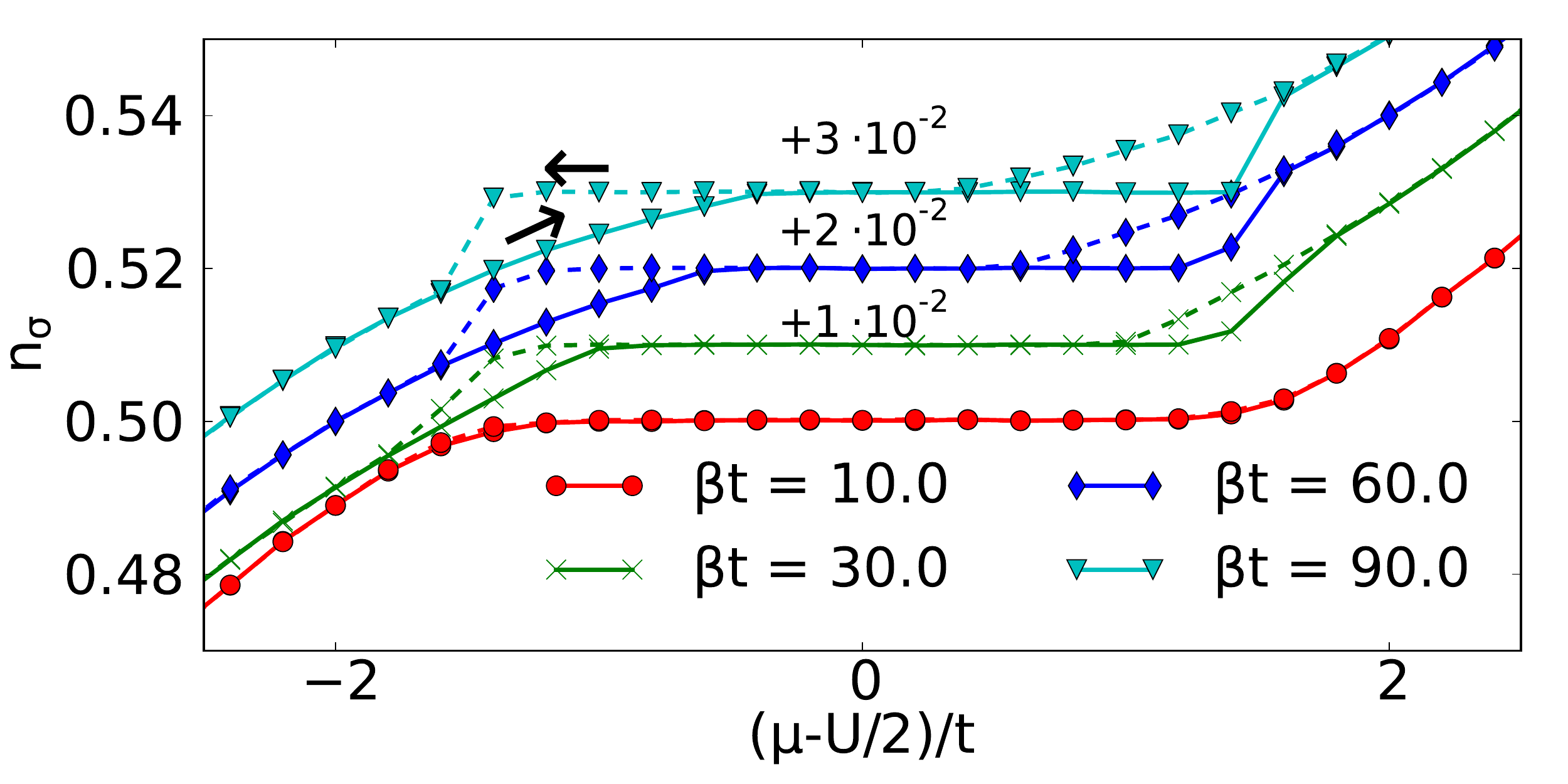}
  \caption[]{(Color online)
    Bulk (3D) electron density as a function of chemical potential for different temperatures at $U/t=13.2$. This value of the interaction is large enough for a gap to open in the spectrum of the solid. A corresponding plateau appears in the $n(\mu)$ curve. The simulations for different $\mu$ values are carried out recursively. Dashed (full) lines correspond to down- (up-) sweep. The hysteresis indicates the coexistence of two solutions. Offsets added for clarity.}
    \label{bulk_den}
\end{figure}

The dependence of on-site density on chemical potential for the Hubbard model on a three-dimensional tight-binding lattice is shown in Fig. \ref{bulk_den} for $U/t = 13.2$. For all temperatures shown there is a plateau in the $n(\mu)$ curve. This is due to the existence of a gap in the density of states (DOS) of the solid at that value of the Hubbard interaction $U$. When the value of the chemical potential is inside the gap, far from the upper and lower Hubbard bands, the charge density does not change noticeably when the chemical potential is varied, due to the vanishing DOS in the gap. As the chemical potential approaches either of the two Hubbard bands, a quasiparticle peak appears at the Fermi level. This redistribution of spectral weight causes a change in the particle denisty and determines the end of the flat region in the $n(\mu)$ curve.

As is well known, there is a region in the $\mu-U-T$ phase diagram of the three-dimensional fermionic Hubbard model in which both a metallic and an insulating phase can exist.\cite{PhysRevB.53.16214,PhysRevB.75.085108} When the values of $\mu$, $U$, and $T$ are such that the system is in this coexistence region, the DMFT converges to either one of the two solutions, depending on the initial guess (seed) for the self-energy. At the value of the Hubbard repulsion that we are considering, $U/t = 13.2$, and when the system is far away from half-filling (when $\mu - U/2$ is large), only the metallic solution exists. If one gradually brings the system closer to half-filling (by varying  $\mu$ in small increments) and uses the converged value of the self-energy at each point as a seed for the subsequent simulation, the system remains on the metallic branch until the metallic solution ceases to exist in the region very close to half-filling. Continuing the recursion beyond half-filling traces out the insulating branch. This is reflected in the hysteresis in Fig. \ref{bulk_den}. Coexistence is possible in the slab as well.\cite{EPJB.8.555} 

\begin{figure}[h!]
    \includegraphics*[width=.5\textwidth]{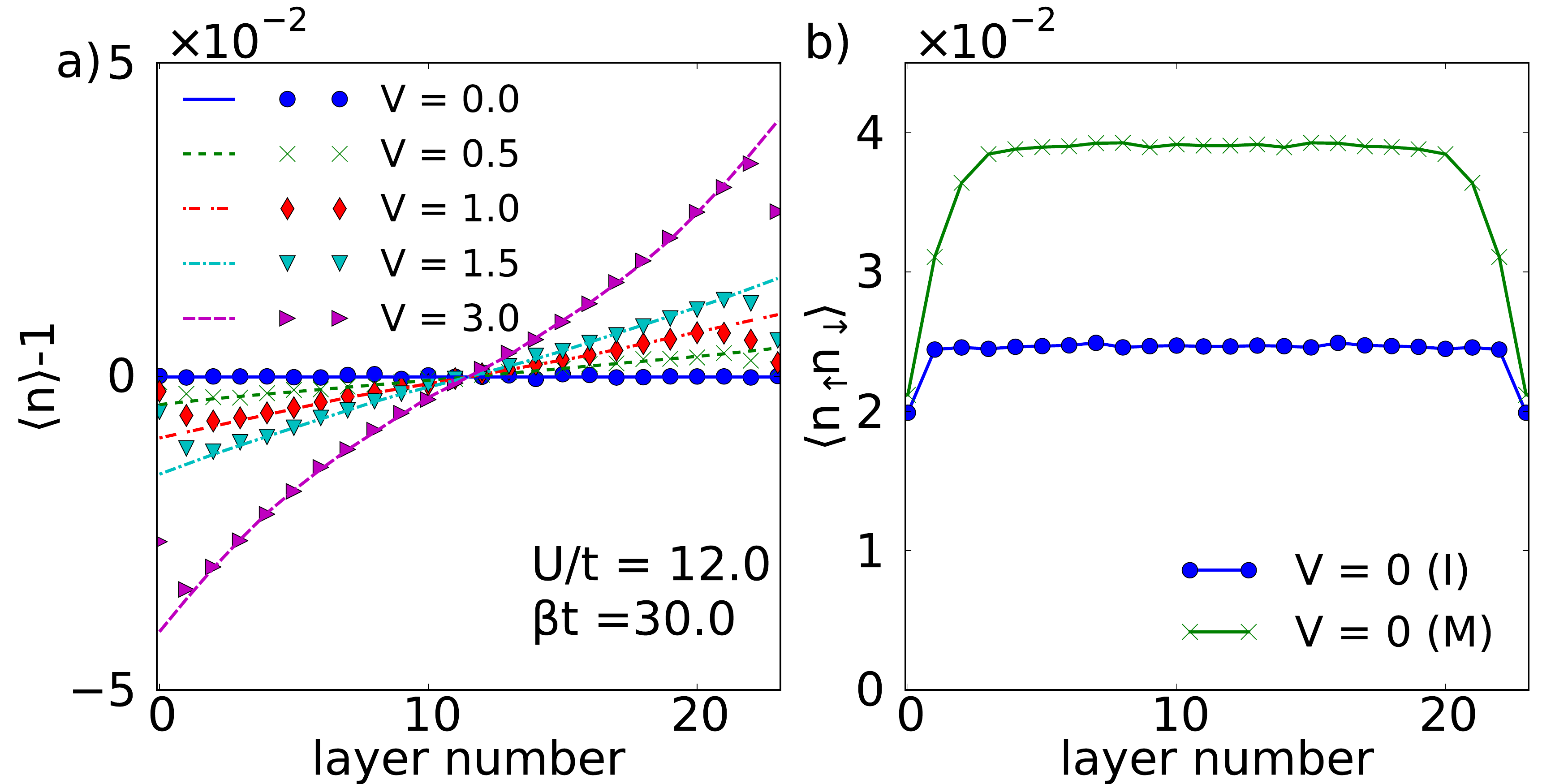}
    \caption[]{(Color online)
Charge density deviation from half-filling across the slab for different values of the applied field. Symbols indicate the full DMFT calculation; lines correspond to the ILA (panel \textbf{a}). Double occupancy across the slab for the insulating and metallic solution at zero field (panel \textbf{b}). $U/t = 12$ and $\beta t = 30$ (both panels).
}
    \label{coexistence_fig}
\end{figure}

In Fig. \ref{coexistence_fig}(a) we show the variation of the charge density of the metallic solution across the slab calculated with DMFT (symbols) and in the ILA (lines) for a number of different values of the applied external field at $\beta = 30$. The interaction strength ($U/t = 12$) is chosen such that the bulk system at half-filling ($\mu = U/2$) would be in the $U-T$ coexistence region, close to its upper boundary, $U_{c2}$, beyond which only the insulating phase is stable. A comparison of the two approaches shows that the DMFT and ILA results are in good agreement in the central parts of the slab, whereas near the surface (within approximately the first four layers) the discrepancy is significant. For larger fields the discrepancy between the DMFT and ILA is significant in a narrower region close to the surface. The main reason for this is the existence of highly-doped regions (close to the surfaces) in which the correlation length is smaller,\cite{PhysRevB.87.035131} which limits the discrepancy between the two approaches to only the surface layer. We expect that the discrepancy between the DMFT and ILA will be larger whenever the correlation length is larger, in particular at very low temperatures and when the first order nature of the transition is weaker. The same trends were observed earlier in calculations based on the Gutzwiller approximation.\cite{PhysRevB.85.085110}

In Fig. \ref{coexistence_fig}(b) we plot the double occupancy, $\langle d_i \rangle = \langle n_{i\uparrow} n_{i\downarrow}\rangle$, across the slab at zero applied external field for both the metallic and insulating solutions. In both the metallic and the insulating case the double occupancy is supressed at the surface. The suppression is due to the surface reduction of the site coordination number, which leads to an enhancement of correlation effects. The surface effect is also visible in the density redistribution in the presence of an electric field (Fig. \ref{coexistence_fig}(a)): the maximum charge deviation does not occur exactly at the surface as is expected from ILA approach. From Fig. \ref{coexistence_fig}(b) it is clear that in the insulating state the double occupancy recovers its bulk value within a single layer, whereas in the metallic case the recovery takes place deeper in the slab (approximately four layers from the surface). This suggests that the correlation length in the metallic state is longer than in the insulating state. The recovery length for the double occupancy at $V = 0$ in the metallic state is related to the extent of the surface effect in the case of applied electric field, which is given by the width of the discrepancy between the full DMFT calculation and the ILA shown in Fig. \ref{coexistence_fig}(a). These two lengths are of approximately the same magnitude.

\subsection{Half-filled case}
\begin{figure}[]
  \includegraphics*[width=.5\textwidth]{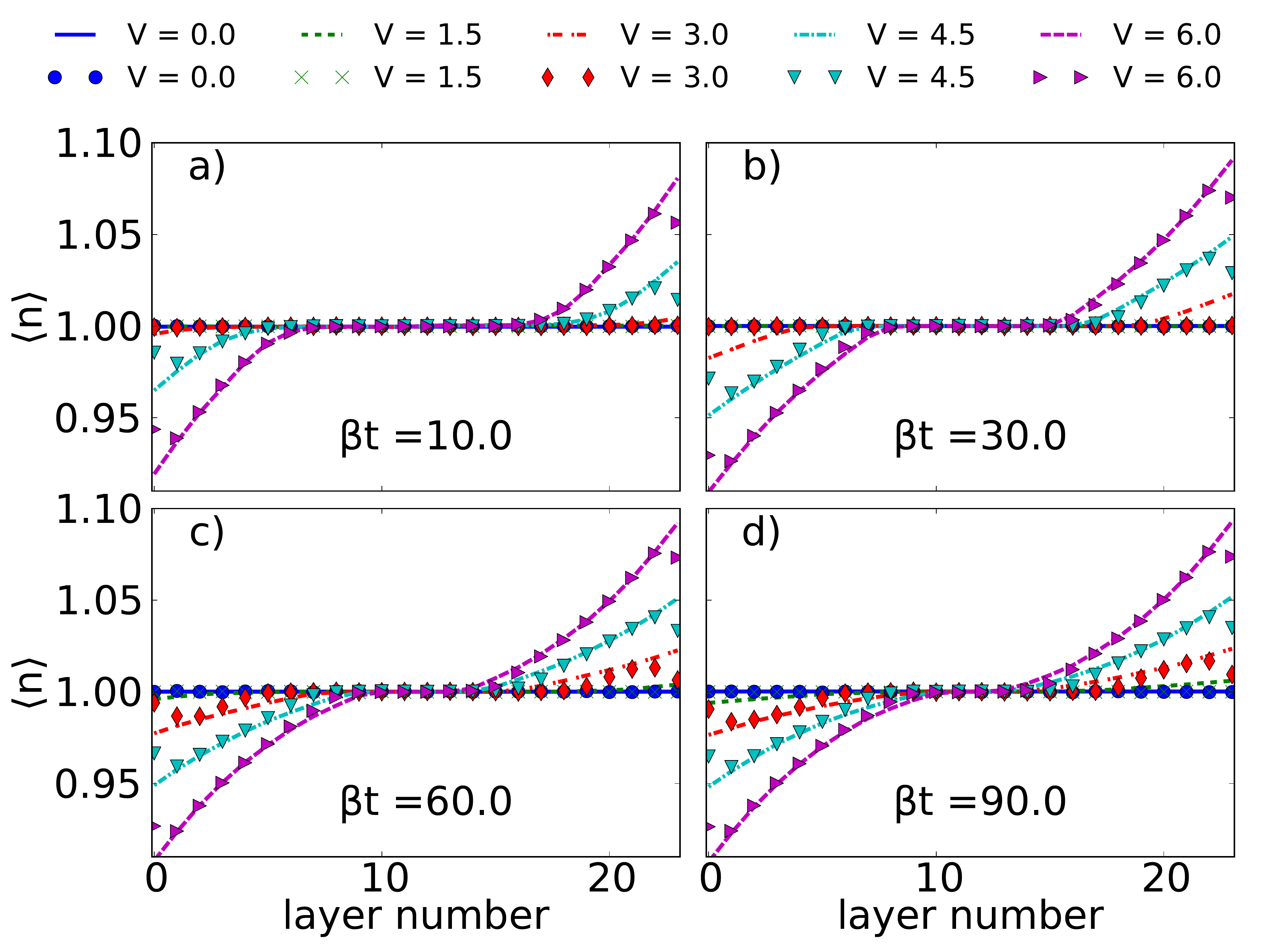}
  \caption[]{(Color online)
    Electron density across the slab at different temperatures and different values of the electric field. The overall density for the slab corresponds to half-filling. Only "metallic" seeds were used. Symbols correspond to the full DMFT calculation, lines correspond to the ILA. $U/t = 13.2$.}
    \label{half_den}
\end{figure}

\begin{figure}[]
  \includegraphics*[width=.5\textwidth]{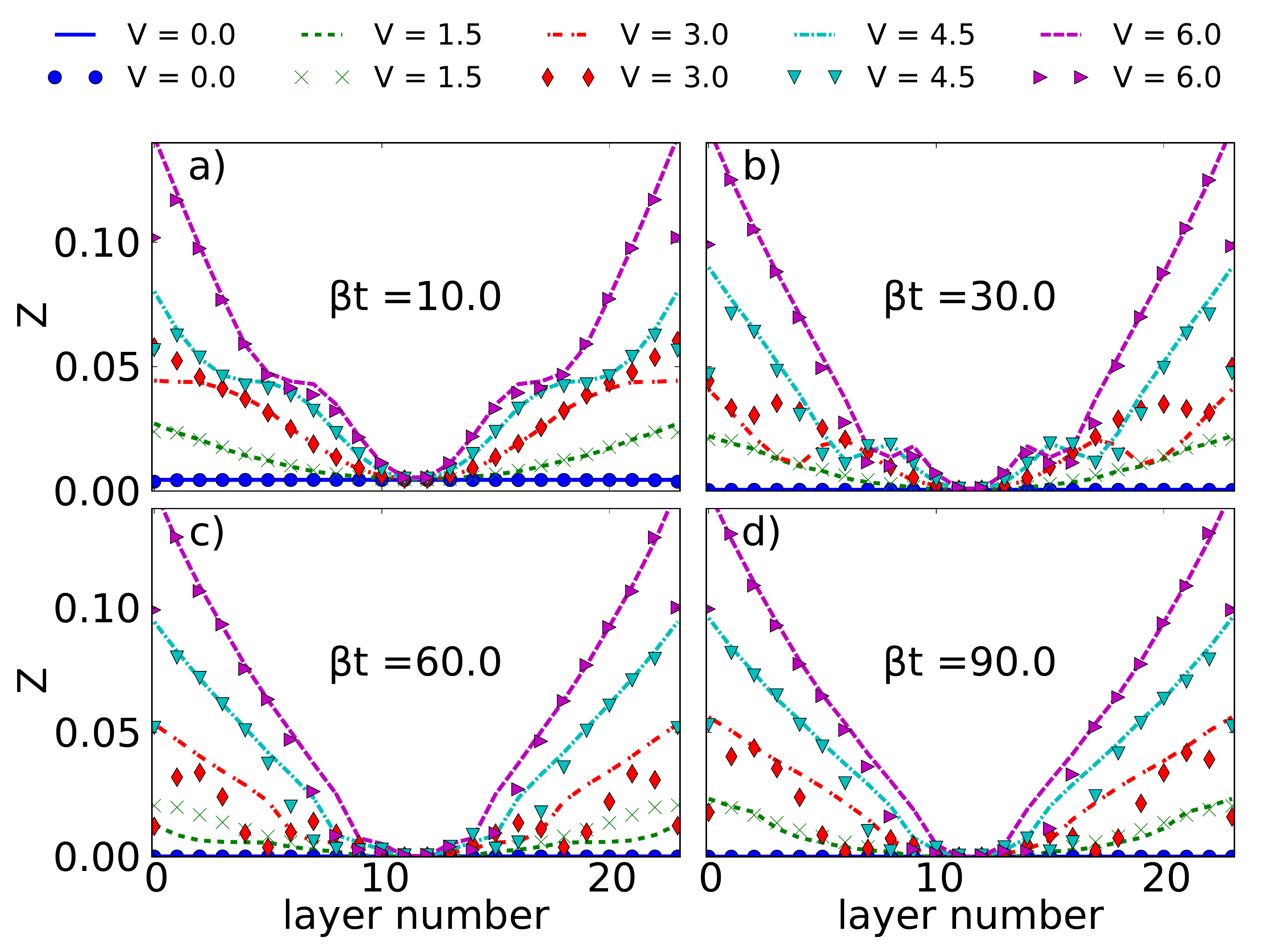}
  \caption[]{(Color online)
  QP weight across the slab for the same simulation parameters as shown in Fig. \ref{half_den}. Symbols correspond to the full DMFT calculation, lines correspond to the ILA.}
    \label{half_den_QP}
\end{figure}

The charge redistribution caused by an electric field applied to a slab at half-filling is shown in Fig. \ref{half_den} ($U/t=13.2$). The value of $U$ is outside the $U-T$ coexistence region for bulk (at $\mu = U/2$) and is such that the slab is insulating at zero field. As discussed in connection with Fig. \ref{bulk_den}, the metallic and insulating phase can coexist in bulk at this value of $U$ away from half-filling. The two phases can coexist in the slab as well, and we find that recursive simulations for different values of $V$ lead to different solutions, i.e. hysteresis (not shown). Whenever two distinct solutions exist, in Fig. \ref{half_den} we show only the metallic one. It is clear from the figure that when the field applied to the slab is below a certain threshold, there is no redistribution of charge in the slab. When the field increases sufficiently, the charge density close to the surfaces of the film starts deviating from half-filling. The deviation is symmetric with respect to the center of the slab, as can be expected from the particle-hole symmetry of the Hubbard model with nearest-neighbor hopping.

The corresponding quasiparticle weights, plotted in Fig. \ref{half_den_QP}, are also enhanced close to the surface in the presence of an electric field. In contrast to what is observed for the density, the QP weight is enhanced close to the surface even at the smallest non-zero value of the applied field which we consider here ($V = 1.5$). In most cases, the highest value of $Z$ in the slab occurs not immediately at the surface but one layer into the slab. Deeper into the slab $Z$ decreases from that maximum value. At the highest temperature in the series ($\beta t = 10$) there is only a narrow central region where the enhancement is not significant. The width of the low-QP-weight region increases as the temperature is lowered.

\begin{figure}[]
  \includegraphics*[width=.5\textwidth]{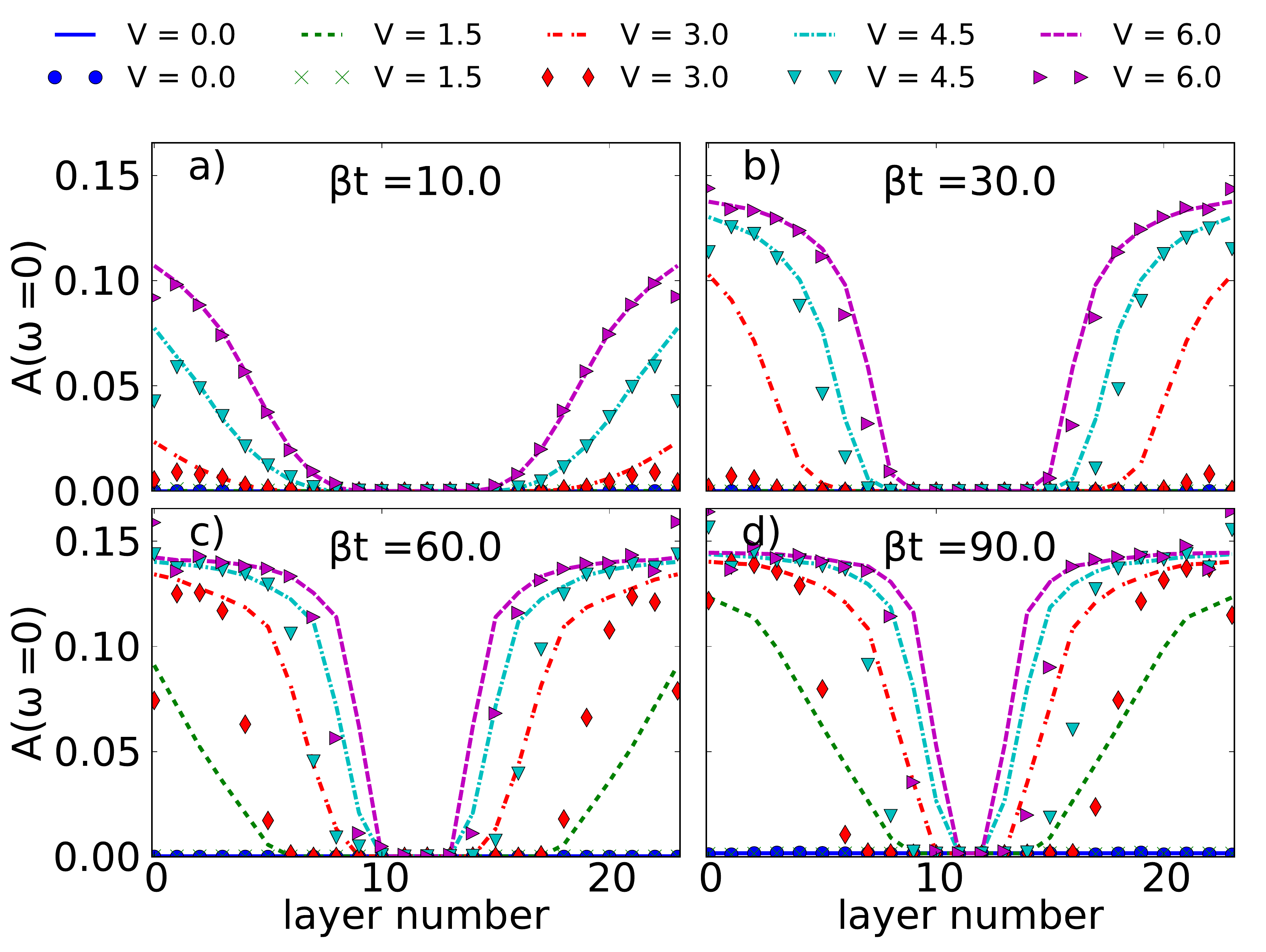}
  \caption[]{(Color online)
    The variation of the DOS at the Fermi level across the slab corresponding to the density and QP weight data shown in Figs. \ref{half_den} and \ref{half_den_QP}. Symbols correspond to the full DMFT calculation, lines corrspond to the ILA.}
    \label{half_den_dos}
\end{figure}

$A(\omega = 0)$, the density of states at the Fermi level, is plotted in Fig. \ref{half_den_dos} for the same simulation parameters as in Figs. \ref{half_den} and \ref{half_den_QP}. Similarly to the QP weight, $A(\omega = 0)$ is enhanced most strongly close to the surfaces. In contrast to what is observed for the QP weight, however, the enhancement of $A(\omega = 0)$ at the lowest non-zero bias ($V=1.5$) is insignificant. Furthermore, except at $\beta t = 10$, at large enough bias, the value of $A(\omega = 0)$ remains nearly constant (close to its maximum value in the slab) in the layers immediately near the surface. The transition to the low-DOS region in the center of the slab is relatively abrupt, compared with what we observe for the QP weight. Contrary to the width of the low-QP-weight region, the width of the low-DOS region decreases with decreasing temperature. Thus, at the highest temperature ($\beta t = 10$) there is a region of high $Z$ and low $A(\omega = 0)$ close to the center of the slab, whereas at lower temperatures there is a region of strongly enhanced $A(\omega = 0)$ and weakly enhanced QP weight.

In combination with the charge redistribution in the slab, the field enhancement of the QP weight and density of states at the Fermi level leads to the formation of conductive channels close to the surfaces of the slab, while the central portion of the slab remains close to its (insulating) $V = 0$ state.

The fact that a certain minimal field is necessary to achieve a "breakdown" of the insulator (i.e. to create an electrically doped region close to the surface in which the QP weight and DOS are significant) can be understood more easily by considering the dependence of density on chemical potential in the 3-dimensional Hubbard model (Fig. \ref{bulk_den}). Due to the presence of a gap in the spectrum for large enough $U$, a plateau analogous to the one in Fig. \ref{bulk_den} appears in the $n$ vs. layer curves in Fig. \ref{half_den}, which corresponds to values of the local chemical potential that are well within the gap. When the externally applied field is large enough, the local value of the chemical potential close to the surface of the slab deviates sufficiently from $U/2$ to induce a change in the local charge density. 

Some of our simulations (not shown) also indicate that when $\beta t \ge 30$ and the applied external field is small, only an insulating solution exists in the slab. In other words, independent of the seed, the simulation converges to the insulating solution throughout the slab. Above a certain value of the field ($V \approx 3.0$), two solutions reappear. This is consistent with what we observe in bulk (Fig. \ref{bulk_den}).

The temperature dependence of the bulk $n(\mu)$ curves also elucidates what is observed in the slab. At $U/t = 13.2$ the range of $\mu$-values for which both a metallic and an insulating solution exist is clearly larger at lower temperature, extending to values of $\mu$ closer to the middle of the gap. 
On the metallic branch (Fig. \ref{bulk_den}), the plateau in $n(\mu)$ shrinks as temperature drops and almost completely disappears at $\beta = 90$. The insulating solution, in contrast, is less temperature sensitive. This temperature dependence of the bulk metallic solution is consistent with what we observe in the slab (see Fig. \ref{half_den}): the minimum field strength required to "break" the insulator decreases as the temperature drops. The width of the insulating region that remains at half-filling in the center of the slab for a given field also decreases as the temperature is lowered.

Similarly to what is observed in bulk, changes in the local value of the chemical potential in individual layers in the slab do not lead to a significant change in density, unless the change is large enough.

\begin{figure}[h!]
    \includegraphics*[width=.5\textwidth]{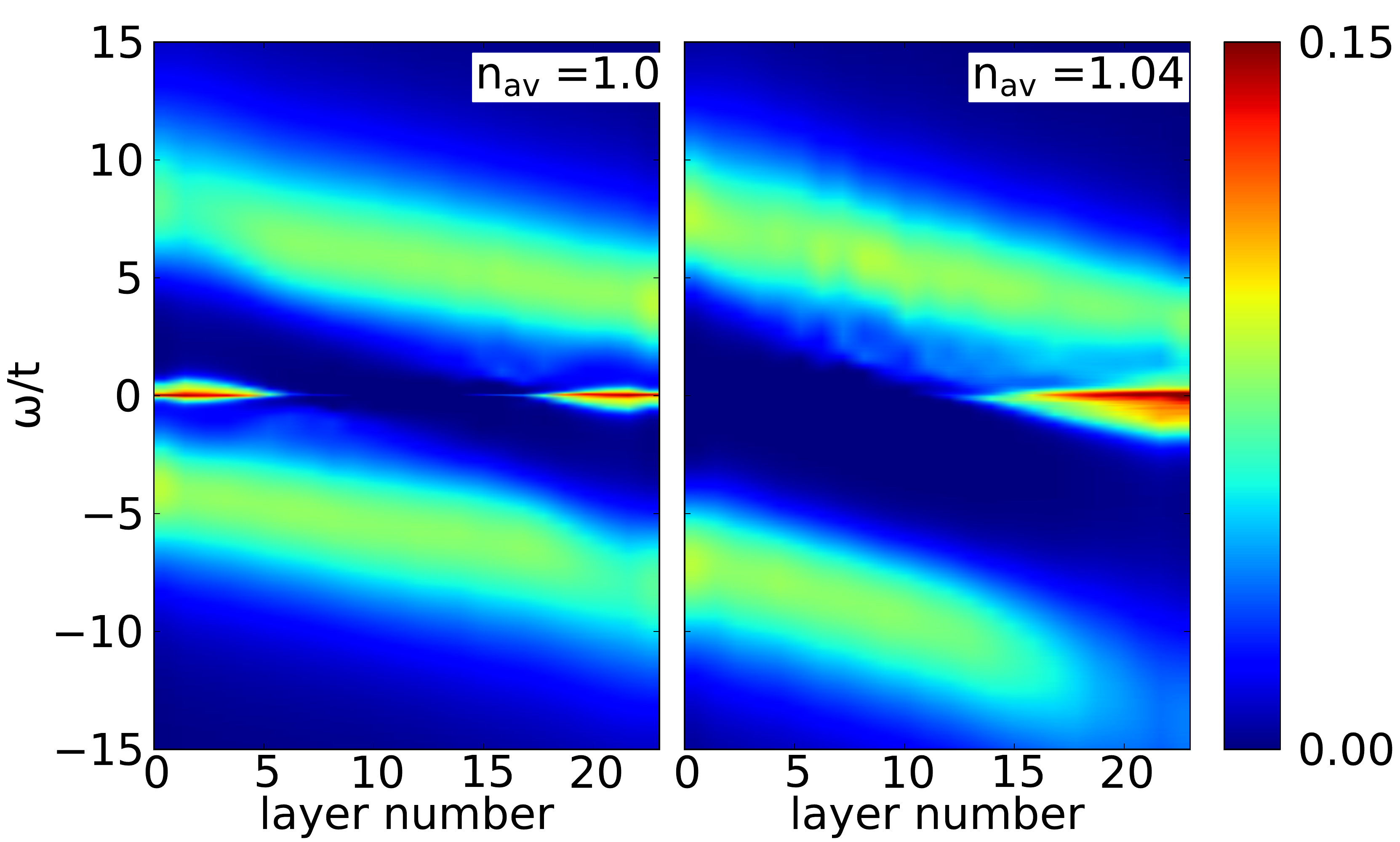}
    \caption[]{(Color online)
    Maximum-entropy method reconstruction of the spectral functions $A_i(\omega)$ throughout the slab. Data is interpolated between layers for clarity. Left panel: $U/t = 13.2, \mu = U/2, \beta t = 30, V = 4.5$ and average density corresponding to half filling. Right panel: $U/t = 16.2, \beta t = 10, V = 6$ and average density per site in the slab $n =1.04$. The MEM results are in very good agreement with those obtained by Eq. (\ref{dos_estimate}).}
    \label{mem_dos_both}
\end{figure}

Maximum-entropy method (MEM) reconstructions\cite{PhysRevB.57.10287,Jarrell1996133} of the real-frequency spectral function $A(\omega)$ provide additional insight into the changes that occur across the slab as bias is increased and parts of the slab become metallic (see the left panel in Fig. \ref{mem_dos_both}). We validated our MEM results in two different ways: comparison with the results of Eq. (\ref{dos_estimate}) reveals very good agreement. Furthermore, calculating the average density in each layer from the real frequency spectral function, according to:
\begin{equation}
n_i = \int_{-\infty}^{+\infty} d\omega A_i(\omega) f(\omega - \mu_i),
\end{equation}
where $A_i(\omega)$ is the MEM spectral function of layer $i$ and $f(\omega)$ is the Fermi-Dirac distribution function, we find excellent agreement with the Monte-Carlo charge density data (Fig. \ref{bulk_den}). The appearance and development of the quasiparticle peaks due to the spectral transfer from the nearby Hubbard bands (the frequency-integrated DOS remains constant) is clearly visible in the figure. The width of the peaks is proportional to the QP weight $Z$ and reaches a maximum approximately one layer into the slab from each of the surfaces, which is consistent with Fig. \ref{half_den_QP}. Similarly, the abrupt rise of $A(\omega = 0)$ is reproduced (in the MEM plot $A(\omega = 0)$ corresponds to the height of the QP peak). The MEM results for the doped case (Fig. \ref{mem_dos_both}, right) are discussed in more detail in the following section.

It is clear (Fig. \ref{half_den}) that the ILA estimate of the density agrees fairly well with the full calculation for almost all the temperatures and field strengths we consider. The difference between the two approaches is greatest close to the surfaces of the slab, where the independent plane approximation tends to overestimate the density and fails to capture the dip evident in the full calculation, and for values of $V$ which are close to the "breakdown" bias of the insulator. The ILA underestimates the value of the breakdown bias. The differences between the ILA estimates for $Z$ and $A(0)$ and the full calculation are more pronounced. The largest discrepancy in $Z$ is observed at $V=3$, the bias value at which the system is closest to the breakdown point. At $V=3$ and $\beta t \ge 30$ the ILA and full DMFT results for $Z$ differ not only near the surface, but through a significant part of the slab. A similar observation is valid for $A(\omega = 0)$, but in addition to the significant disagreement between the ILA and DMFT estimates of $A(\omega = 0)$ at $V=3$, the ILA also significantly overestimates the value of $A(\omega=0)$ at $V=1.5$ and $\beta t = 60, 90$. The agreement is better at larger values of $V$ and at higher temperature (lower $\beta$).

\subsection{Screening}
\begin{figure}[]
  \includegraphics*[width=.5\textwidth]{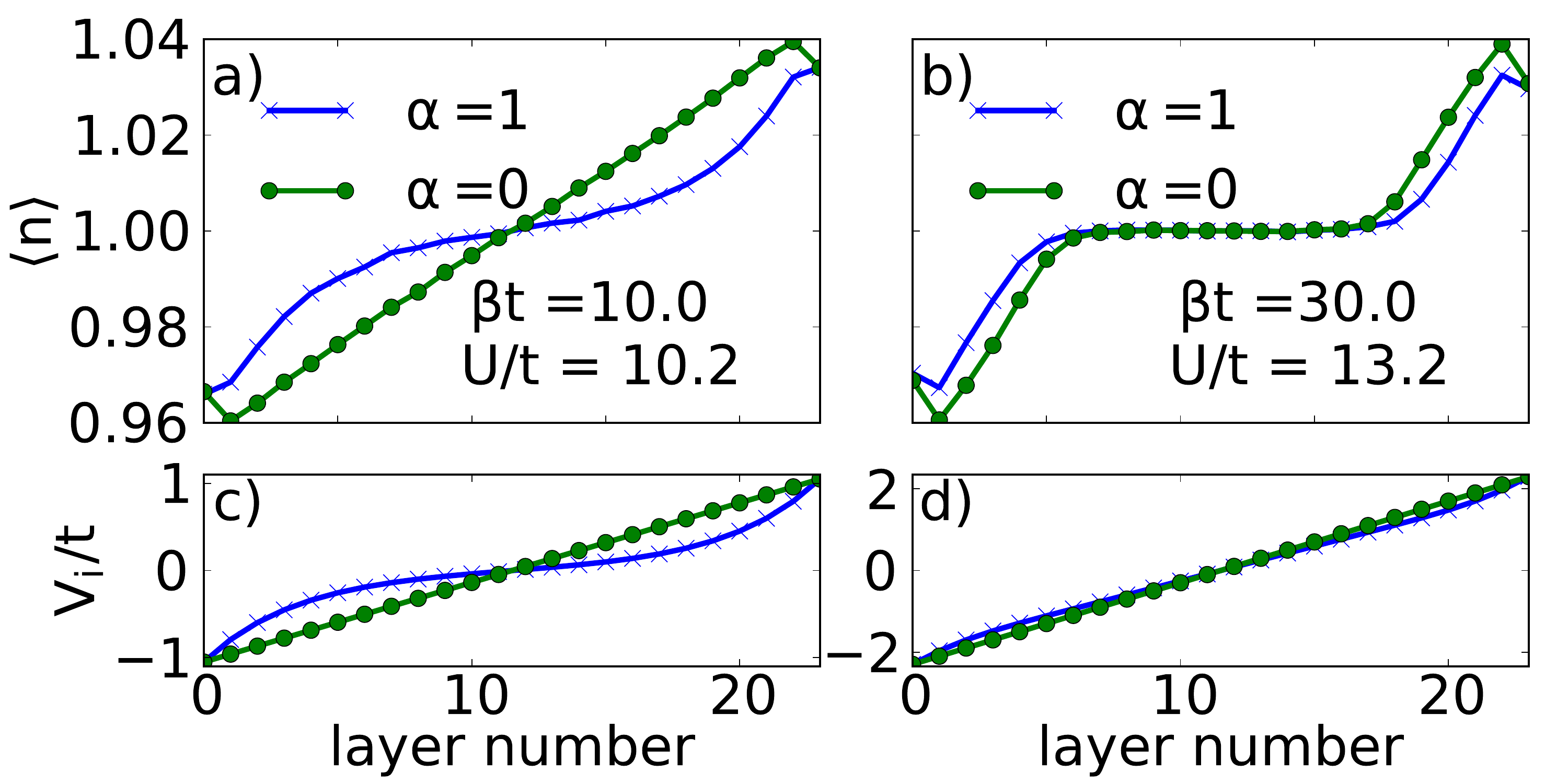}
  \caption[]{(Color online)
    The effect of screening on a metallic slab, panels \textbf{a}, \textbf{c}, and an insulating slab, panels \textbf{b}, \textbf{d}. The effect is much more pronounced for the metallic case. $\alpha$ is the screening parameter described in Sec. \ref{mnm}.}
    \label{screened}
\end{figure}

We also investigate the effect of screening on the half-filled slab (see Fig. \ref{screened}). In Figs. \ref{screened}(a,c) we show the density and potential across a slab in the metallic phase ($U/t = 10$ and $\beta t = 10$) in the presence of screening and in the unscreened case. For the screened case we use $\alpha = 1$. The total potential drop across the slab in both cases is the same (this means that in the screened case the applied external field $V$ is greater). In practice, we first fix the externally applied field $V$ and the screening strength $\alpha$ and allow the simulation to converge to a self-consistent solution that satisfies Eq. (\ref{coulomb}). This leads to a reduced potential drop across the slab. Then a non-screened simulation with the same potential drop is performed for comparison. The effect of screening is to reduce the penetration of the field in the slab. As a consequence, the density deviation (from half-filling) is reduced, especially in the central region of the slab. Qualitatively the results remain very similar: the deviation from half-filling is symmetric with respect to the central region and largest close to the two surfaces in the slab. Screening masks the influence of the surface on the density: in the absence of screening the maximum of the denisty does not occur at the surface but one layer into the slab. In contrast, in the screened case, the maximum is at the surface. The surface reduction effect is still visible, but softened, due to the decreased penetration of the field in the slab. The effect of screening on an insulating slab ($U/t = 13.2$, $\beta t = 30$) is analogous, Figs. \ref{screened}(b,d). In contrast to the metallic case, however, the effect is much less pronounced.

\subsection{Doped case}
\begin{figure}[h!]
    \includegraphics*[width=.5\textwidth]{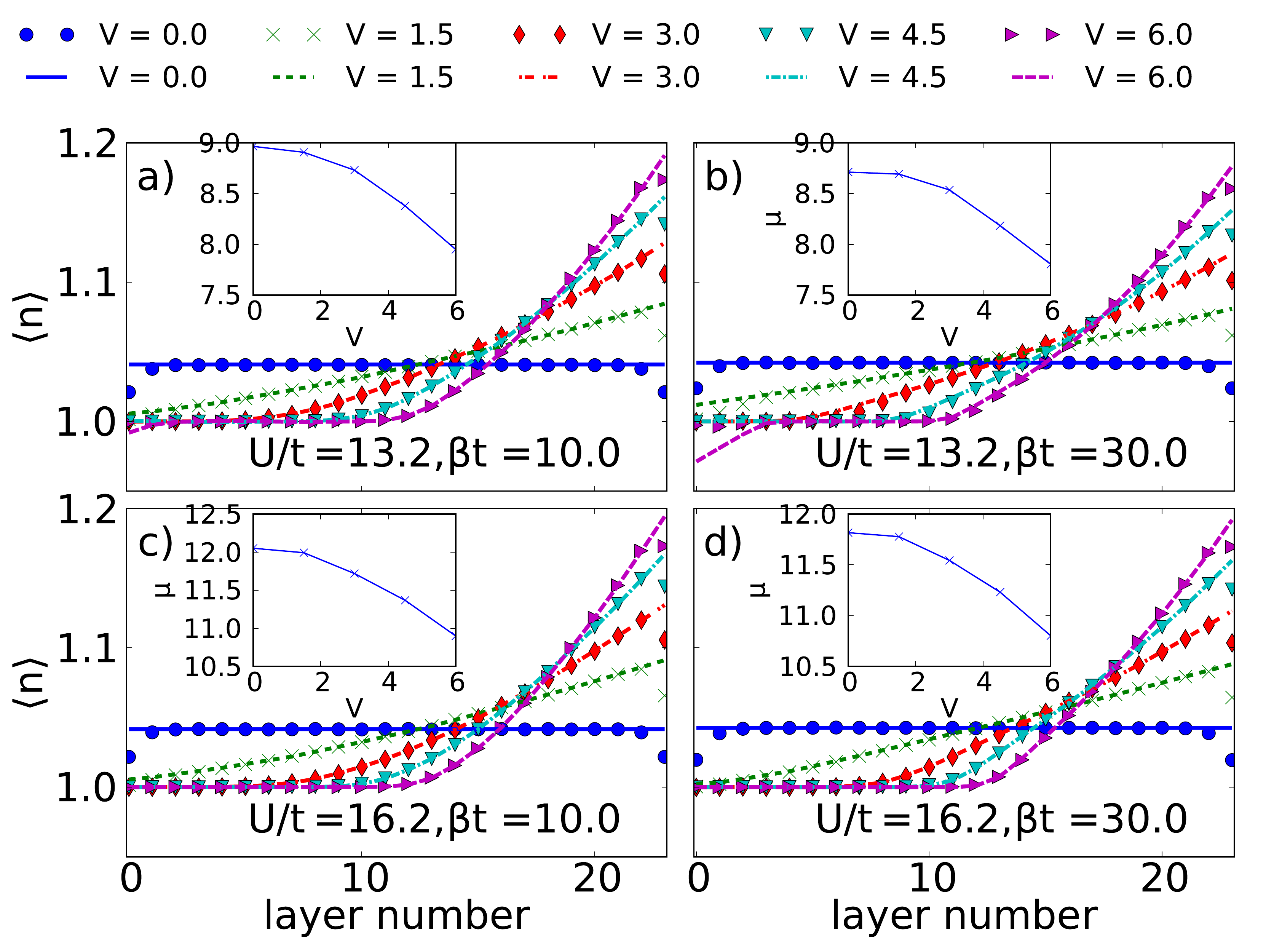}
    \caption[]{(Color online)
    Variation of the electron density throughout a 24 layer slab for selected values of the electric field, Hubbard $U$ and temperature. The average density for the whole slab is fixed at 1.04 electrons per site. The insets show how the chemical potential of the slab changes as the field is increased. This adjustment is necessary to keep the average electron density in the slab fixed.}
    \label{const_den}
\end{figure}

\begin{figure}[h!]
    \includegraphics*[width=.5\textwidth]{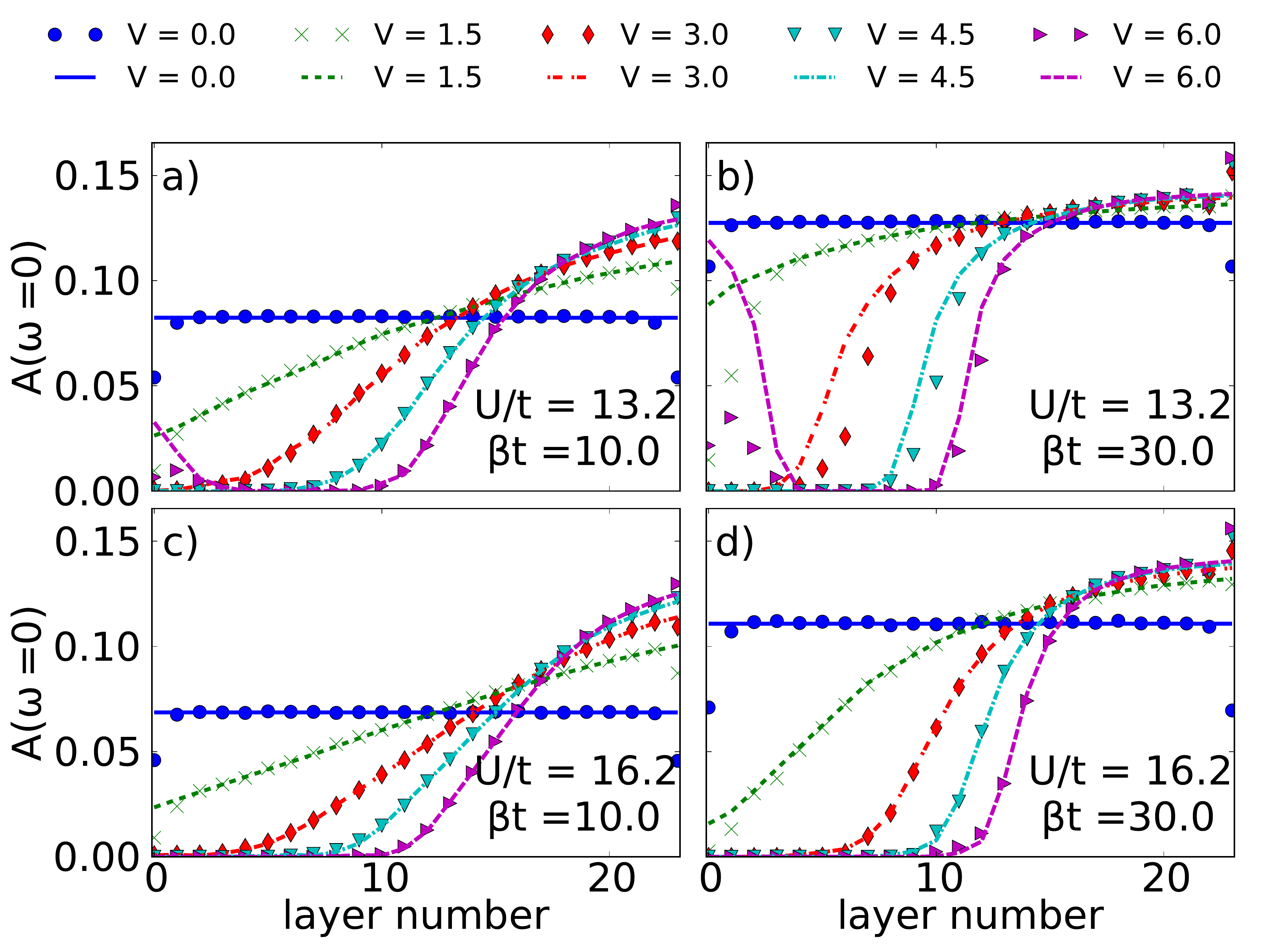}
    \caption[]{(Color online)
    Variation of the density of states at the Fermi level across a 24 layer slab for the same values of the electric field, Hubbard $U$ and temperature as in Fig. \ref{const_den}. The average density for the whole slab is fixed at 1.04 electrons per site. Symbols correspond to the full DMFT calculation, lines corrspond to the ILA.}
    \label{const_den_dos}
\end{figure}

\begin{figure}[h!]
    \includegraphics*[width=.5\textwidth]{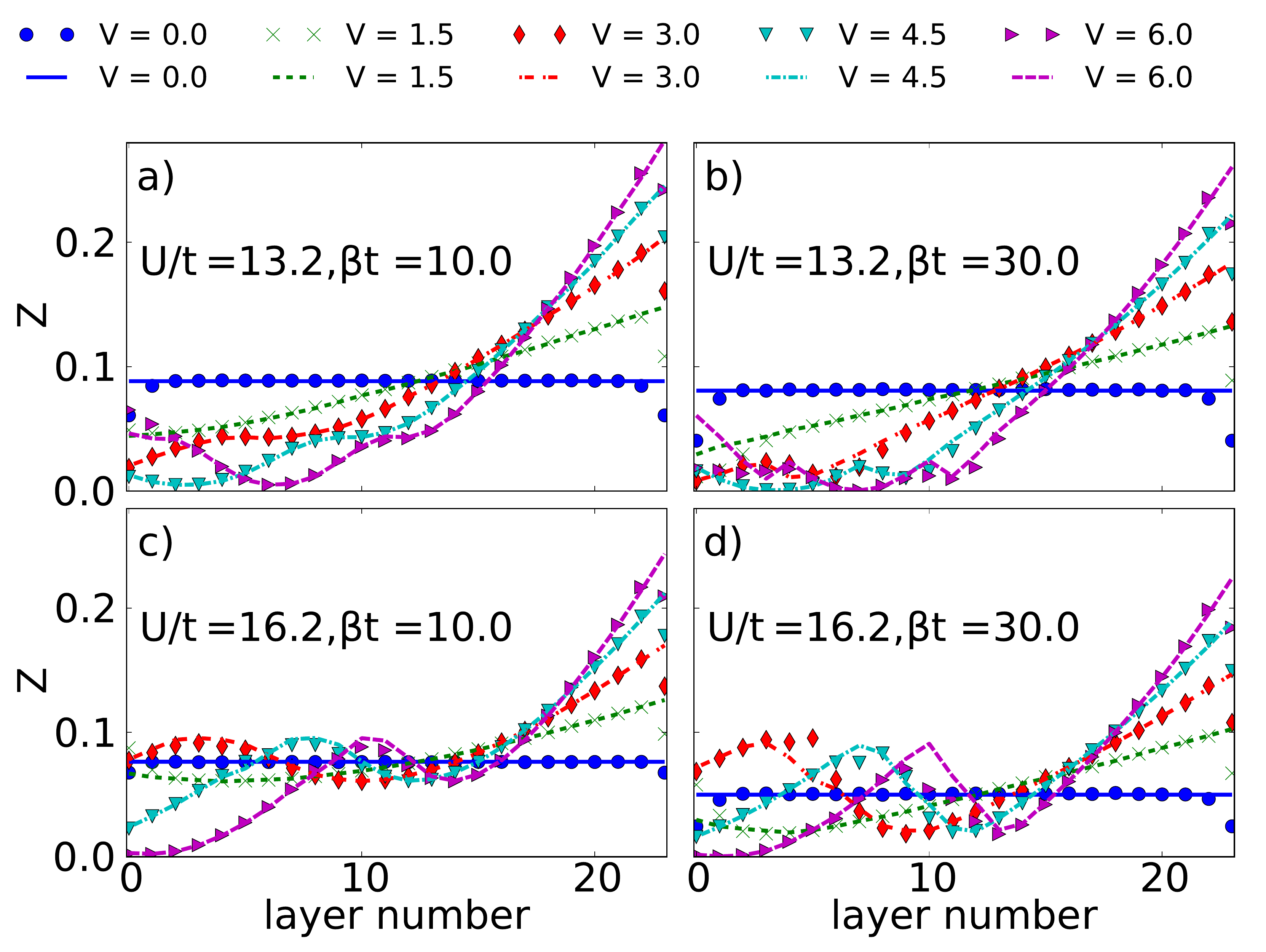}
    \caption[]{(Color online)
    Variation of the QP weight across a 24 layer slab for the same values of the electric field, Hubbard $U$ and temperature as in Figs. \ref{const_den} and \ref{const_den_dos}. The average density for the whole slab is fixed at 1.04 electrons per site. Symbols indicate the full DMFT calculation; lines correspond to the ILA. }
    \label{const_den_QP}
\end{figure}

The effect of the field on a doped slab is quite different. In Figs. \ref{const_den} - \ref{const_den_QP} we show the charge density redistribution (DMFT and ILA results), $A(\omega = 0 )$, and quasiparticle weight across the slab for four different values of $U$ and $\beta$ ($U/t = 13.2, 16.2$; $\beta t = 10, 30$). At our chosen doping ($n = 1.04$) and in the absence of a field ($V = 0$), both the density of states at the Fermi level (Fig. \ref{const_den_dos}) and the QP weight (Fig. \ref{const_den_QP}) are significant throughout the slab in all four cases, indicating a metallic phase. In this case even a small field causes charge redistribution. Since the average density in the slab is kept fixed, the pile-up of charge on one side is accompanied by a decrease in the charge density on the other side. However, if the value of $U$ is large enough (i.e. sufficient for a gap to appear in the DOS of the solid at half-filling), this decrease in the charge density on the depleted side pauses as soon as half-filling is reached. From that point on, as the magnitude of $V$ increases the charge build-up on the other side is compensated by an increase in the width of the half-filling region. This is what we observe here for $U/t = 13.2$ and $U/t = 16.2$. Thus, at large $U$, as the field increases one of the sides of the slab becomes more conducting while an insulating layer of increasing thickness develops at the opposite end of the slab. Note that the density redistribution is accompanied by a change in the QP weight: the QP weight is enhanced on the side with excess charge and suppressed on the "depleted" side. The density of states at the Fermi level follows a similar trend. As the temperature is lowered from $\beta t = 10$ to $\beta t = 30$ the transition between the low- and high-conductivity regions becomes more abrupt (Fig. \ref{const_den_dos}). Variation of $U/t$ in the range considered ($13.2 - 16.2$) does not significantly influence the density results. The effect of increasing $U$ on $A(\omega = 0)$ is most pronounced at $V=0$, in which case it leads to a decrease of the DOS at the Fermi level throughout the slab. 

The insets in Fig. \ref{const_den} show how the chemical potential should be adjusted in order to keep the average amount of charge per layer in the slab constant (at $n = 1.04$) at each bias. The adjustment is necessary due to the non-linear dependence of charge density on $\mu$ (see Fig.~\ref{bulk_den}). Experimentally, this situation corresponds to a slab that is electrically insulated from its environment (the overall amount of charge in the slab remains constant). The observed reduction of chemical potential in the presence of external electric field indicates that if connected to a charge reservoir (e.g. via metallic leads) an overdoped system would tend to absorb charge when an electric field is applied. In simulations with doped slabs at constant $\mu$ we do observe an increase in the amount of charge in the slab as the field increases (not shown). This charge absorption is cuased by the nonlinearity of the $n(\mu)$ curve and not by a potential difference between the system and its environment.

At higher values of the interaction ($U/t= 16.2$) oscillations appear in the QP weight vs. layer number. The effect of lowering temperature is to make the oscillations more pronounced (cf. Fig. \ref{const_den_QP}, lower two panels). In the presence of a significant DOS, these oscillations would correspond to alternating regions of low and high mobility of the quasiparticles in the film. However, since $A(\omega = 0)$ is rather low in the part of the film where these oscillations occur (see Fig. \ref{const_den_dos}), in practice the variation in conductivity would be small. The fact that these oscillations in $Z$ are also reproduced by the ILA indicates that they are not related to the slab geometry.

It is worth noting that different parts of the slab can coexist in different states, i.e. that the metallicity of one part of the slab does not penetrate throughout the slab to destroy the low-QP-weight and low-DOS region in other parts of the slab.
The MEM reconstruction of the slab spectral functions is shown in the right panel of Fig. \ref{mem_dos_both} for $U/t = 16.2$, $\beta t = 10$, and $V = 6$. In this case the agreement with Eq. (\ref{dos_estimate}) and DMFT density data is again very good. Due to the doping, the position of the lower and upper Hubbard bands with respect to the Fermi level is not symmetric, in contrast to the half-filled case. This asymmetry is also reflected in Fig. \ref{const_den} to Fig. \ref{const_den_QP}. From the MEM reconstruction it is clear that once again the QP and DOS enhancement on the right side of Figs. \ref{const_den_QP} and \ref{const_den_dos} are associated with the appearance of a QP peak. On the basis of the MEM reconstruction, it is to be expected that at sufficiently strong fields the lower Hubbard band will also approach the Fermi level and a quasiparticle peak will appear on the low-doped side as well, which will cause a "secondary breakdown" to occur at the insulating side of the film.

The doping and the resulting asymmetry lead to a very different response compared to the half-filled case. In the latter case, the application of the field to the insulating slab leads to the formation of symmetric zones of relatively high conductivity close to the surfaces, whereas in the doped case, which is metallic in the absence of a field, application of a field causes one side to become more conductive, whereas the other side becomes insulating, until the field is large enough for the "secondary" breakdown to occur.

The ILA results for the charge density match the full calculation very nearly in almost all cases, except in the surface layers, and - when the bias is in the vicinity of the "secondary breakdown" - in a narrow region close to the surface on the insulating side. The agreement on the overdoped side improves as the bias increases. We attribute this to the short correlation length in the highly-doped (metallic) regime. On the underdoped side the discrepancy is most significant at low temperature ($\beta t \ge 30$), close to the "secondary breakdown".  As in the half-filled case, the ILA underestimates the breakdown voltage. This tendency to overestimate the extent of the metallic phase is reflected also in the ILA results for the density of states, Fig. \ref{const_den_dos}, and QP weight, Fig. \ref{const_den_QP}. Overall, the agreement between the DMFT and ILA is much better for the doped case than for the half-filled slab, which can be partially attributed to the fact that at most biases and temperatures a large part of the slab is metallic.

\begin{figure}[h!]
    \includegraphics*[width=.48\textwidth]{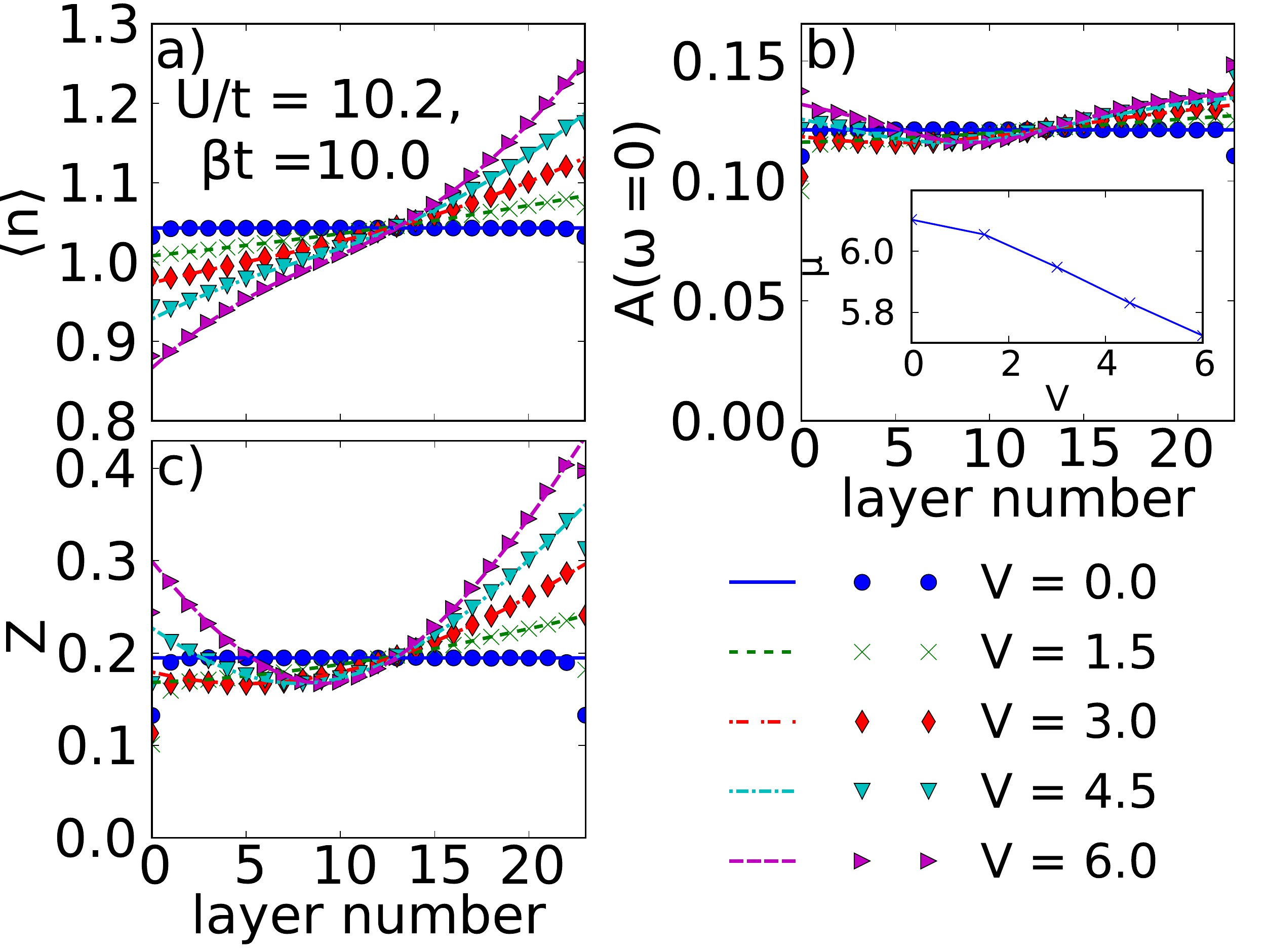}
    \caption[]{(Color online)
    Electron density, DOS at the Fermi level, and QP weight across a doped 24 layer slab for the same values of the electric field as in Figs. \ref{const_den} - \ref{const_den_QP}, but lower $U$. The average density in the slab is fixed at 1.04 electrons per site. The inset shows the change of $\mu$ necessary to keep the average density constant for different values of $V$. Symbols correspond to the full DMFT calculation, lines corrspond to the ILA.}
    \label{const_den_metallic}
\end{figure}

When the Hubbard interaction is lowered to $U/t=10.2$ (Fig. \ref{const_den_metallic}), no half-filling region of significant width develops, as there is no plateau in the $n(\mu)$ curve at this interaction strength. At this value of $U$ the QP weight exhibits a minimum as a function of layer number, and is enhanced close to the slab surfaces. Expectedly, as in the large $U$ case, even a weak field is sufficient to cause a redistribution of charge. The agreement between the DMFT and the ILA results for density, $A(\omega = 0)$, and QP weight is excellent at all bias values we consider. The largest discrepancy between the two approaches occurs in the two layers immediately at the surface and is most significant in the value of $Z$ (panel \textbf{c}).

\section{Conclusion}
\label{conc}
In summary, we investigate the properties of strongly correlated thin films under bias using IDMFT and examine the validity of a computationally cheaper approximation. We observe switching behavior in both half-filled and doped films. In the half-filled slab a sufficiently strong field (larger than some threshold) is necessary to produce conducting regions near the surfaces of the film. For doped films, there is no threshold field and the application of a field initially causes one side of the film to become more insulating, before a secondary breakdown occurs. Taking screening into account does not lead to qualitative changes of the results. We therefore expect that the conclusions for the non-screened case will remain valid in the presence of screening. The hysteretic behavior associated with the first order Mott transition observed in bulk persists in the slab, and should lead to memory effects in devices.

In spite of the breakdown of useful concepts from band theory, such as band bending, the local independent layer approximation accurately reproduces the full IDMFT calculations in both the half-filled and doped cases, except in the layers immediately at the surface of the slab and close to transition points. Our calculations confirm and extend earlier findings in theoretical studies of colossal magnetoresistance, where DMRG calculations in 1D systems indicated the existence of a universal density-potential relation of the interface Mott transition.\cite{PhysRevLett.95.266403} Thus the ILA may be used as a fairly reliable and quick first estimate for the charge density (and to a lesser extent for the quasiparticle weight and DOS at the Fermi level). This may allow for calculations of device properties such as e.g. electrostatic charge distribution, switching behavior, differential capacitance, which are not easily accessible with full DMFT calculations due to the larger computational cost.

\section{Ackonwledgements}
This work was partially funded by the Flemish Fund for Scientific Research (FWO - Vlaanderen) under FWO grant G.0520.10 and by the SITOGA FP7 project. Most of the calculations were performed on KU Leuven's \mbox{ThinKing} HPC cluster.

\bibliographystyle{apsrev4-1}
\bibliography{references.bib}

\end{document}